\documentclass[pdflatex,sn-nature, iicol]{sn-jnl}

\usepackage{graphicx}%
\usepackage{multirow}%
\usepackage{amsmath,amssymb,amsfonts}%
\usepackage{amsthm}%
\usepackage{mathrsfs}%
\usepackage[title]{appendix}%
\usepackage{xcolor}%
\usepackage{textcomp}%
\usepackage{manyfoot}%
\usepackage{booktabs}%
\usepackage{algorithm}%
\usepackage{algorithmicx}%
\usepackage{algpseudocode}%
\usepackage{listings}%
\usepackage{stfloats}%
\usepackage{tabularx}%
\usepackage{threeparttable}%
\usepackage{makecell}
\usepackage{booktabs}
\usepackage{mathtools}
\usepackage{wasysym}

\theoremstyle{thmstyleone}%
\theoremstyle{thmstyletwo}%

\theoremstyle{thmstylethree}%

\makeatletter
\providecommand{\authorcr}{\par}
\newcommand{\authorskip}[1][\baselineskip]{\par\addvspace{#1}}
\makeatother

\begin{document}

\title[Article Title]{Nano Machine Intelligence: From a Communication Perspective}


\author[1,2]{\fnm{Sangjun} \sur{Hwang}}\email{sangjun848@yonsei.ac.kr}
\equalcont{These authors contributed equally to this work.}

\author[3]{\fnm{Bon-Hong} \sur{Koo}}\email{bh9.koo@samsung.com}
\equalcont{These authors contributed equally to this work. \authorskip \textit{Invited Paper} }

\author[1]{\fnm{Ho Joong} \sur{Kim}}\email{meerschaum@yonsei.ac.kr}

\author[1]{\authorcr \fnm{Jang-Yeon} \sur{Kwon}}\email{jangyeon@yonsei.ac.kr}

\author*[1]{\fnm{Chan-Byoung} \sur{Chae}}\email{cbchae@yonsei.ac.kr}


\affil*[1]{\orgdiv{School of Integrated Technology}, \orgname{Yonsei University}, \orgaddress{\city{Seoul}, \postcode{03722}, \country{South Korea}}}

\affil*[2]{\orgdiv{BK21 Graduate Program in Intelligent Semiconductor Technology}}

\affil*[3]{\orgdiv{Network Business}, \orgname{Samsung Electronics}, \orgaddress{\city{Suwon}, \postcode{16677}, \country{South Korea}}}




\abstract{We present an AI-integrated molecular communication link validated on a benchtop nanomachine testbed representative of subdermal implants. The system employs an indium–gallium–zinc-oxide electrolyte-gated FET (IGZO-EGFET) functionalized with glucose oxidase as a biocompatible receiver, a microfluidic channel with a syringe-pump transmitter using on–off keying (OOK), and a machine-intelligence pipeline that addresses model mismatch and hardware non-idealities. The pipeline integrates: (i) a modular universal decoder robust to vibration-induced noise, chemical delay, and single-tap intersymbol interference; (ii) a lightweight pilot-only synchronizer that estimates symbol intervals; and (iii) a virtual-response generator that augments data and scales symbol duration. Experiments across multiple chips and sessions demonstrate end-to-end chemical text transmission with consistent error-rate reductions compared to na\"{\i}ve thresholding and standard neural baselines. By coupling biocompatible hardware with learning-based detection and generative augmentation, this work establishes a practical route toward AI-native nanomachine networks and higher rate molecular links, while providing a system blueprint adaptable to other biochemical modalities.}




\maketitle

\section{Introduction}\label{sec1}
Molecular communication is an emerging paradigm that uses molecules to conveys information instead of electromagnetic waves, enabling communication at biological and nanoscale dimensions~\cite{Akyildiz2008Nanonetworks,Farsad2016Survey}. It is a promising alternative in environments where radio performance is degraded, such as intrabody medical implants~\cite{Lee2023jcn}. A typical system operates by mapping digital information to prescribed molecular concentrations, which are then released into a defined channel. At the destination, a receiver measures the molecular signal and converts it back into digital form. However, due to dispersion and related transport effects, these channels exhibit inherent randomness~\cite{Kadloor2009Frame}. Their complex and nonlinear dynamics of molecular communication channels make closed-form modeling difficult, motivating the use of artificial intelligence (AI) methods to improve end-to-end performance~\cite{BartunikKirchnerKeszocze}.

Traditionally, physical-layer techniques in molecular communication, including channel modeling, detection, and coding, have relied on statistical signal processing and probabilistic models~\cite{Nakano2005}. For instance, in diffusion-based channels, detector design and capacity analysis have been addressed using statistical mechanics and information theory, but these methods are only effective when tractable mathematical models exist~\cite{koo2021}. While model-based designs can approach theoretical optima, they often miss real-world complexities and can suffer from model mismatch~\cite{llatser2014n3sim}.

To bridge this gap, early studies introduced benchtop testbeds that physically transmit information using molecular signals. Subsequent work has closed the loop by implementing theoretical designs in hardware, demonstrating that improved models lead to enhanced prototype performance~\cite{farsad2013tabletop, koo2016molecular}. Focusing on in-vessel applications, researchers have developed pipe such as channels and microfluidic environments that emulate blood vessels~\cite{rogers2016parallel, Nariman2017Vessel}. Furthermore, prototypes using electromagnetically controlled nanoparticles have been demonstrated, suggesting the feasibility of control schemes for in-body applications~\cite{unterweger2018experimental}.

Recent work in molecular communication increasingly pairs theoretical analysis with experimental testbeds, recognizing the frequent divergence between theory and real-world operation~\cite{Farsad2014jsac, hofmann2024}. This approach is particularly critical for in-body communication, where biological variability and environmental constraints are difficult to model analytically. To address the complexities found in these experimental settings, data-driven approaches, such as machine learning and deep learning, can approximate nonlinear relationships from data and achieve near-optimal operation without explicit channel knowledge~\cite{koo2020deep}.

\subsection{Related works}
Deriving closed-form channel models for molecular communication is inherently difficult due to complex fluid dynamics~\cite{Yilmaz2014}, partial absorption at the receiver, multi-antenna coupling, and the effects of advection and turbulence. As a result, data-driven methods have become prevalent. These methods typically involve training machine learning models on simulation data or limited experimental measurements to approximate the channel response, then integrating the detection stage~\cite{Yilmaz2017MLSignal}. For example, in systems with a spherical transmitter and an absorbing receiver, obtaining a closed-form expression for the number of received molecules is challenging. Therefore, supervised neural networks have been used to learn the cumulative number of arrivals as a function of distance, diffusion coefficient, time, and related parameters~\cite{TorresGomez2023XAI}. These models have been incorporated directly into demodulation and detection pipelines, extended to MIMO configurations, and adapted to systems with partially absorbing receivers. In such systems, convolutional and recurrent neural networks can jointly estimate cumulative arrivals and perform detection within a single processing chain~\cite{Baydas2023}.

On the receiver side, diffusion and reaction induce long memory, strong intersymbol interference (ISI), time variability, and imperfect synchronization~\cite{hyun2024}. These factors make the assumption of an exact channel model or perfect channel state information (CSI) is unrealistic~\cite{Lee2017spawc}. A more effective strategy is to minimize channel parameter estimation and instead learn decision boundaries directly from data. This concept has been realized with sequence level detectors that map an input sequence to an output sequence while leveraging temporal context. A representative example is the sliding bidirectional recurrent neural network (Bi-RNN), which applies a fixed-length window that moves along the stream and uses a bidirectional recurrent network to classify the symbol at the center of each window. Trained on both synthetic Poisson model data and experimental datasets, this method enables real-time streaming detection without an explicit channel model~\cite{Farsad2018tsc}.

More recently, transformer-based detectors have been proposed to address the limitations of conventional RNNs and convolutional neural network (CNN) in capturing long-range dependencies. In one study, a fast matrix-based particle simulator was combined with the detector to generate large labeled datasets. This approach yielded a reliable detection pipeline under severe ISI while keeping the training cost at a practical level~\cite{lu2023mcformer}. Furthermore, several studies have employed machine learning to model molecular arrival signals over MIMO channels~\cite{Changmin2015Testbed}. These data-driven models are trained and validated on experimental data collected from diverse testbeds, including both benchtop and microfluidic platforms. Results consistently show that these models capture complex effects, such as inter-link coupling and detector nonlinearities, more accurately than conventional parametric models~\cite{Lee2020}.

Building on these advances, many studies now focus on training complete entire end-to-end pipelines for molecular communication. One representative approach jointly models the transmitter and receiver as an autoencoder and augments the transmitter with deep reinforcement learning~\cite{Junejo2021AutoencoderMCN}. This enables the joint optimization of the modulation mapping and the decision boundary without an explicit channel model~\cite{Mohamed2019}. In parallel, synchronization on hardware testbeds implementing theoretical designs has been actively explored to correct transmission errors observed in practical experiments. For example, one study demonstrates a reinforcement learning based receiver that adapts its compensation parameters online to reduce missed detections and timing misalignment on a large-scale testbed~\cite{Debus2023RLReceiver}. In another example on a pH-modulation testbed, physically interpretable features such as polynomial approximation coefficients and piecewise derivatives are combined with neural networks. Using pretraining with synthetic and augmented data and a confidence-based ensemble, this detector achieved over $99.9\%$ symbol accuracy~\cite{Vakilipoor2022Hybrid}.

\begin{table*}[!t]
\centering
\caption{Comparison to prior works.}
\label{tab1}
\begin{tabular*}{\textwidth}{@{\extracolsep{\fill}}lcccc@{}}
\toprule
\textbf{Study} & \makecell[l]{\textbf{System-level}\\ \textbf{operation}} & \textbf{Synchronization} & \textbf{Experiment} & \textbf{Data generation} \\
\midrule
Proposed & {\Large \(\Circle\)} & {\Large \(\Circle\)} & {\Large \(\Circle\)} & {\Large \(\Circle\)} \\
Baydas (2023)~\cite{Baydas2023} & {\Large \(\times\)} & {\Large \(\times\)} & {\Large \(\times\)} & {\Large \(\times\)} \\
Luo (2024)~\cite{luo2024integrating} & {\Large \(\times\)} & {\Large \(\times\)} & {\Large \(\Circle\)} & {\Large \(\times\)} \\
Lu (2023)~\cite{lu2023mcformer} & {\Large \(\times\)} & {\Large \(\times\)} & {\Large \(\times\)} & {\Large \(\Circle\)} \\
Vakilipoor (2022)~\cite{Vakilipoor2022Hybrid} & {\Large \(\times\)} & {\Large \(\times\)} & {\Large \(\Circle\)} & {\Large \(\times\)} \\
Farsad (2018)~\cite{Farsad2018tsc} & {\Large \(\times\)} & {\Large \(\times\)} & {\Large \(\Circle\)} & {\Large \(\times\)} \\
Mohamed (2019)~\cite{Mohamed2019} & {\Large \(\times\)} & {\Large \(\times\)} & {\Large \(\times\)} & {\Large \(\times\)} \\
Debus (2023)~\cite{Debus2023RLReceiver} & {\Large \(\Circle\)} & {\Large \(\Circle\)} & {\Large \(\Circle\)} & {\Large \(\times\)} \\
\bottomrule
\end{tabular*}
\end{table*}

\subsection{Novelty}
This paper focuses on nanomachines that can communicate in realistic settings. We develop a biocompatible receiver operating in a microfluidic environment and construct an experimental chain that integrates receiver learning, online operation, and data generation. Guided by the physical and chemical error sources observed in practice, we design a modular pipeline that addresses the main communication tasks. We encode physical hypotheses into the module definitions, and we train each module with data-driven learners using multiple branches within each module. Building on this foundation, we propose a universal decoder that operates across modules. To relieve data bottlenecks, we introduce a virtual response generator that synthesizes response time series and supports environment oriented augmentation, including symbol-interval scaling. We also develop a lightweight synchronization method that identifies the symbol interval using only pilot responses, which reduces dependence on prior channel knowledge and improves system-level operability.

As summarized in Table.~\ref{tab1}, our work differs from prior studies in several key areas. Our approach is motivated by the challenges of practical deployments, where mechanical errors, chemical degradation, and data synchronization issues interact. Under these complex conditions, designs developed mainly through simulation are difficult to implement and often fail to transfer to hardware.

\begin{itemize}
    \item \textbf{Biocompatible nanomachine design:} Development of an IGZO-EGFET glucose sensor integrated into a microfluidic environment, enabling realistic subdermal implant scenarios.
    \item \textbf{Modular machine-intelligence pipeline:} A universal decoder with dedicated modules for vibration-induced noise, chemical delay compensation, and single-tap ISI mitigation.
    \item \textbf{Data augmentation via virtual response generation:} A novel generator that synthesizes realistic molecular response signals, alleviating data scarcity and enabling symbol-interval scaling.
    \item \textbf{Lightweight synchronization:} A pilot-only synchronizer that infers symbol intervals without prior channel knowledge, improving system operability on hardware.
    \item \textbf{Experimental validation:} End-to-end demonstration of chemical text transmission with consistently reduced BER compared to na\"{\i}ve thresholding and standard ANN/RNN baselines.
\end{itemize}

\begin{figure*}[t!]
	\centering
	\includegraphics[width=0.99\linewidth]{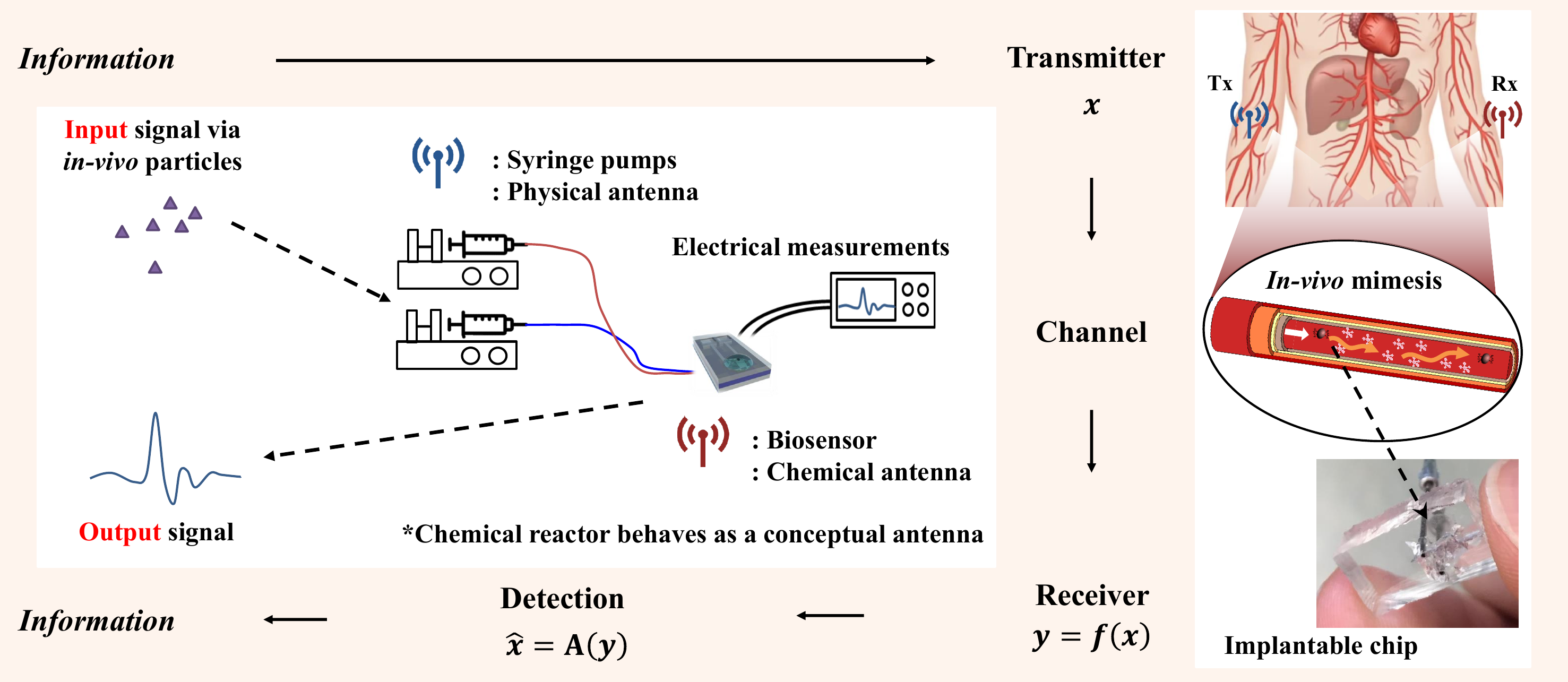}
	\caption{An illustration of the proposed system concept, where a subdermal biosensor detects information encoded in the concentration of glucose molecules.}
	\label{Fig.Concept_art}
    \vspace{2pt}
    \noindent\rule{\linewidth}{0.4pt}
\end{figure*}

This work experimentally validates and extends the theory of AI-aided molecular communication by implementing a testbed that is usable in realistic settings. Our system model targets a subdermal implant scenario. As shown in Fig.~\ref{Fig.Concept_art}, we constructed a nanoscale communication testbed that models links between nanodevices inside the human body. In the experiments, information is modulated onto glucose molecules that propagate through a tube-shaped conduit. The receiver is a custom sensor that we fabricated and embedded in a chunk of diaphanous protein.

Grounded in theoretical analysis and experimental observation, we identify errors that arise when transitioning the system from theory to practice. We address these challenges with machine intelligence to present a more robust solution. We participated in all fabrication steps for the sensor and the testbed. Small process variations cannot be fully controlled, and they create non-ideal effects. Our proposed method compensates for these effects and demonstrates practical effectiveness through successful end-to-end data transmission. Unlike prior work, our approach is fundamentally grounded in experimental evidence, using machine learning to achieve technical validity confirmed by measurements.

The remainder of the paper is organized as follows. Section~\ref{sec2} introduces the nano transmitter and receiver, the complete testbed, along with a simple baseline detector. Section~\ref{sec3} presents a reference communication model that explains the observed non-linearity and includes the simple detection rule. Section~\ref{sec4} details the machine-intelligence-based method. Section~\ref{sec5} establishes a communication protocol using the learned receiver and demonstrates the transmission of an example text message. Finally, Sections~\ref{sec6} and~\ref{sec7} report the experimental results and conclusions.

\section{Nanomachine design}\label{sec2}
In this work, we present the design and experimental validation of a self-contained molecular communication system based on nanomachines. Unlike conventional molecular communication platforms that rely on external instrumentation, our architecture embeds both sensing and transmission capabilities directly within nanoscale devices. Key challenges in achieving such nanomachines include biocompatibility, scalable fabrication, and end-to-end communication performance. This section first identifies these constraints and presents our strategies to overcome them. Then details the fabrication workflow, describes the experimental testbed, and summarizes the resulting performance metrics.
	
	\subsection{Nano transducer design: IGZO-based EGFET}
	\label{Sec4.chip}

To address the challenges outlined above, we employ an indium–gallium–zinc oxide electrolyte-gated field-effect transistor (IGZO-EGFET). IGZO-based devices have garnered significant interest in biomedical electronics because both semiconductors and their by-products are considered safe for \textit{in-vivo} use \cite{Hwang2016}. Compared to alternative materials, such as graphene or polysilicon nanowires, nanosheets—IGZO-EGFETs offer superior field-effect mobility and on/off current ratio, which result in reduced noise and higher signal fidelity \cite{Shen2014Biosensor, Huang2015Biosensor}. Moreover, their low-voltage operation and intrinsic label-free sensing capability make them ideal for overcoming biocompatibility and efficiency barriers.

For proof of concept, we functionalize the EGFET surface with glucose oxidase (GOx), transforming it into a glucose sensor. Following the mechanism reported in \cite{Sohn1997ISFET}
\begin{equation}
    \begin{aligned}
    \mathrm{C_6H_{12}O_6 + H_2O + O_2 \xrightarrow{GOx}} \\
    \mathrm{C_6H_{11}O_7^- + H_2O_2 + H^+},
    \label{eq_glucose}
    \end{aligned}
\end{equation}
the oxidation of glucose induces a local pH change. This change produces a surface potential variation that modulates the transistor channel current. During calibration, the device exhibited a sensitivity of 52.8$\pm$1.3 mV per pH, a response time under 5 seconds, and a linear detection range of 0.05 to 2 mM.

		\begin{figure}[!t]
		\centerline{\resizebox{1.0\columnwidth}{!}{\includegraphics{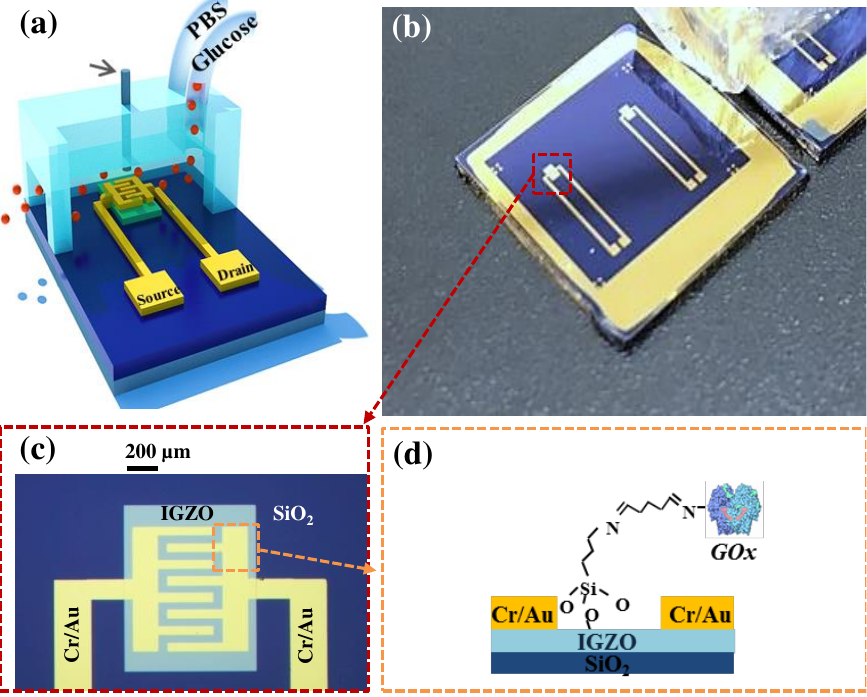}}}
		\caption{The conceptual diagram of the fabrications of IGZO-based EGFET. (a) An environment of the chip measuring chamber consists of three-pin measures from gate, source, and drain. (b) A physical sample of the fabricated chip. (c) A conceptual top view shows the pattern of IGZO-based EGFET fabrication~\cite{Juan2014IGZO}. (d) A front view shows the immobilizations of GOx on IGZO so that the chemical reaction with injected glucose molecules may occur.}
		\label{Fig.Chip_Fabrication}
        \noindent\rule{\linewidth}{0.4pt}
	\end{figure}
    
	The production of the sensor is briefly depicted in Fig.~\ref{Fig.Chip_Fabrication}. In detail, the process has two steps: the fabrication of IGZO-based EGFETs and GOx immobilization on the EGFETs.
	
	\subsubsection{IGZO-based EGFET fabrication}
	The IGZO thin film was deposited at a thickness of 50 nm by radio frequency (RF) sputtering on 100 nm of thermally-grown SiO$_2$/highly-doped Si substrate. The molar ratio of the IGZO sputter target was In$_2$O$_3$:Ga$_2$O$_3$:ZnO = 1:1:1. The sputtering power density was 2.46 W/$\mathrm{cm}^{2}$ with a working pressure of 5 mTorr. The partial pressure of $\mathrm{O_{2}}$ in a mixture of Ar gas was approximately 1.6\% to obtain the semiconducting properties of IGZO. After the deposition of IGZO, Cr/Au of 10/100 nm thickness was deposited by sputtering as the source and drain electrodes of the FETs. The device was annealed in air at 300 $^\circ{C}$ for an hour to reduce the contact resistance between the electrodes and IGZO. An Ag/AgCl leak-free reference electrode (Warner) was used as the gate electrode. The devices were operated under phosphate buffered saline (1X PBS, pH: 7.4, Sigma-Aldrich) for glucose detection. The 1X PBS electrolyte is suitable for preparing a glucose solution, so that the osmolality and ionic concentrations of the solution match those of the human body. Finally, a polydimethylsiloxane (PDMS) chamber was attached to the top of the FETs to contain the electrolyte.
	
	\subsubsection{GOx immobilization}
	We immobilized 10 mg/ml of GOx (from Aspergillus niger, Sigma-Aldrich) on the surface of IGZO and used it as an enzyme for identifying glucose. Initially, $\mathrm{O_{2}}$ plasma was treated on the surface of IGZO. For silanization, the hydroxylated IGZO device was immersed for three hours in a 5$\%$ solution of (3-Aminopropyl) tri-ethoxysilane (APTES, Sigma-Aldrich) diluted with ethanol. In order to remove the weakly-bonded APTES molecules from the IGZO surface, it was rinsed with DI water and dried. Glutaraldehyde (Sigma-Aldrich) diluted with DI water to form a $1\%$ solution was treated to crosslink the functional amine group between the surface of the silanized IGZO and the GOx molecules. Then, the GOx solution was drop-casted onto the IGZO films, dried for two hours, and rinsed again.
	
	\begin{figure*}[t]
		\centerline{\resizebox{1.7\columnwidth}{!}{\includegraphics{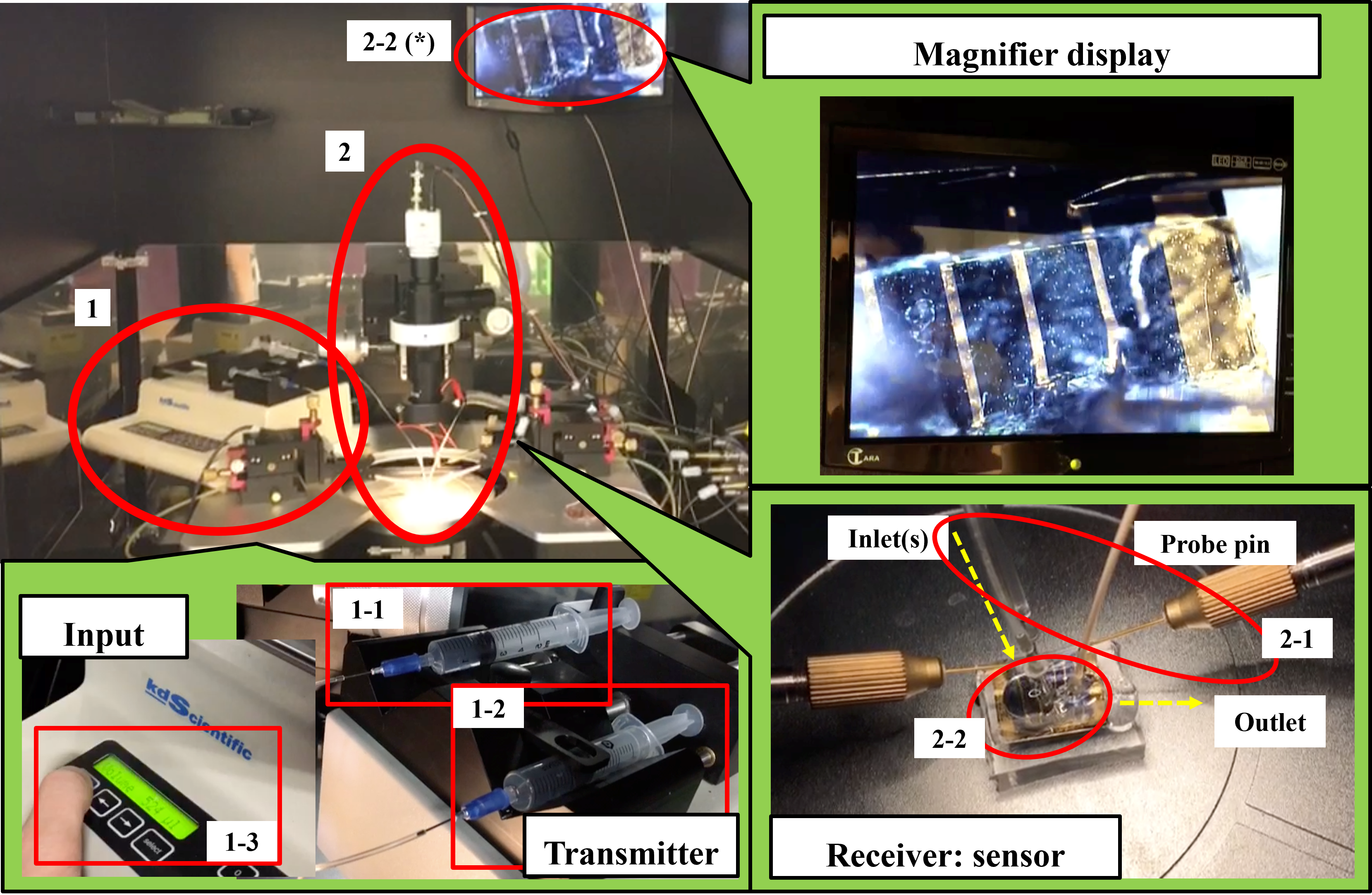}}}
		\caption{Overview of the experimental testbed for molecular communication. The setup is composed of two primary sections: (1) a transmitter, which includes a glucose syringe (1-1), a saline syringe (1-2), and a pump controller (1-3); and (2) a receiver system, featuring a probing chamber (2-1) and the custom-fabricated biosensor under observation (2-2).}    
		\label{Fig.Testbed}
        \noindent\rule{\linewidth}{0.4pt}
	\end{figure*}
	
	\subsection{Experimental testbed}
	\label{Sec4.testbed}

Our experimental platform (Fig. \ref{Fig.Testbed}) consists of three main modules~\cite{Changmin2015Testbed}: a transmitter, a microfluidic channel and a receiver equipped with auxiliary pumps, a probing chamber, and a real-time signal visualizer.

\begin{itemize}
	\item \textbf{Transmitter :} A syringe pump (kdScientific LEGATO 100) dispenses glucose solution with $0.5\%$ accuracy in volume and timing. The pump connects to the receiver via silicone tubing whose inner diameter mimics that of a small human vein. To mitigate mechanical artifacts, we insert fixed guard intervals between symbol injections and cap the data rate at 2 bps.
	\item \textbf{Channel:} The tubing segment emulates \textit{in-vivo} flow. Its length and the background flow rate, supplied by an auxiliary peristaltic pump, are independently adjustable variables in our experiments. Additional side-injection ports enable co-flowing buffer streams to better replicate physiological conditions at junctions.
	\item \textbf{Receiver:} The receiver system employs on–off keying (OOK) modulation, where a binary '1' corresponds to a glucose pulse, and a ‘0’ to no injection within each time slot. The probing chamber monitors the local concentration waveform, and extracts the dominant frequency component (up to 8 Hz) in real time.
\end{itemize}
    
	Table.~\ref{Tab.Experiments_parameters} summarizes the parameter settings across different data rate scenarios. For each experimental run, we transmitted between 50 and 100 symbols, manually tuning the drift and injection rates to maintain baseline stability. Guard intervals reduce residual accumulation and bit error probability. In total, we analyzed 2,000 transmitted bits, with the detailed results presented in the following section.	
	
\begin{table}[!t]
\centering
\caption{The set of parameters and choice of values.}
\label{Tab.Experiments_parameters}
\begin{tabular}{lll}
		{Notation} &  {Term} & {Value, Variations} \\
		\hline
		$t_s$      & Symbol interval    &  0.5, 1, 2, 3 (s) \\
		$t_w$   & Injecting duration &  0.25, 0.5, 1, 2 (s)\\
		$v$     & Drift rate         &  0.5, 1 (ml/min)\\
		$r$     & Glucose injection rate          &  1, 2, 4 (ml/min)\\
\end{tabular}
\end{table}


\section{Baseline system model}\label{sec3}
Building on the testbed described in Sec.~\ref{sec2}, this section details the practical limitations encountered during our experiments. We then present the methods developed to overcome these challenges and enhance system performance. From an implementation standpoint, the major difficulties fall into two broad categories: chemical instability inherent to the fabrication process and mechanical errors that arise from the sensor's limited sensitivity.



	
	\subsection{Nonlinearity observations}
	\label{Sec4.Nonlinearity}
    The IGZO implemented for EGFET operation undergoes multiple process steps during fabrication. During film formation, oxygen vacancy defects frequently appear in the microstructure, resulting in an amorphous structure that impedes charge transport and introduces errors in the precise output of sensor~\cite{Noh2011OxygenVacancy}.
    
Moreover, the sensing reaction shown in ~\eqref{eq_glucose} is irreversible. The GOx enzyme irreversibly produces $H^{+}$, which alters the local pH and induces both physical and chemical defects at the receiver. A primary cause of instability is the consumption of the immobilized GOx enzyme on the sensor surface. During continuous communication, this irreversible reaction leads to a changing baseline pH, causing variations in the response of the sensor. In addition to this instability, the receiver chip lacks an internal power supply and relies on non-rechargeable batteries, which shortens the lifetime of sensor. Omitting a power module keeps the device compact and inexpensive for mass production, which aligns with the practical strategy of periodically replacing the receiver. Nevertheless, even within a single experimental setup, the measured responses vary across chips due to chemical changes and other environmental factors.

The sensors and communication devices in the receiver are highly sensitive to vibration and are therefore enclosed within a protective chamber. However, the junction between the receiver and the intermediate rubber pipe becomes clogged with chunk of diaphanous protein, creating bottlenecks that disturb the fluid flow. This non‑ideal flow converts system vibrations into spike‑like noise, which the sensor interprets as physical noise. Such behavior is expected not only in the testbed but also in \textit{in-vivo} or other application scenarios.

Beyond this unavoidable physical noise, many communication systems assume perfect time synchronization during simulation. While this assumption can be relaxed for high‑speed carriers such as electromagnetic waves, it becomes critical in slow propagation media like molecular communications. Accordingly, intersymbol time synchronization delays are observed in the present testbed. When the transmitter is manually operated to produce constant emission values, the release and reception moments within each time slot do not align perfectly. One common mitigation technique is to enlarge the time slot and use a convolutional sliding window. However, this approach is limited because it requires a known impulse response and linear ISI~\cite{Noel2017Async}.

\subsection{Simple correction algorithm}
	\label{Sec4.simple_correction_algorithm}

To cope with the unpredictable chemical and mechanical errors observed in our testbed, we propose a machine learning-based solution. We hypothesize that by training on experimental data, which inherently includes these nonlinearities, we can develop a robust detection model suitable for real-world deployment. This approach is expected to compensate for both the chemical instabilities near the sensor and the various mechanical errors.

To apply machine learning within the testbed, we use pilot signals to tune various errors. Based on this, we heuristically design three detection algorithms: \textit{na\"{\i}ve successive thresholding}, \textit{spike classifying}, and a \textit{vanilla artificial neural network (ANN)}. Given a transmitted binary sequence \textbf{x} and a received sequence \textbf{y}, each algorithm seeks a mapping function $f$ that minimizes the Hamming distance between the original and decoded sequences. Since the system is asynchronous, the received sequence \textbf{y} is divided bit‑wise where each \textbf{y}$_i$ corresponds to $\hat{\textbf{x}}_i$ while including idle and reactive intervals. Using the pilot signal as reference, $\textbf{y}_1$ is taken as the origin, and each bit is delineated with a time slot of duration $t_s$. With the probing frequency set to 8Hz, each $\textbf{y}_i$ therefore contains $8t_s$ elements.

\begin{figure}[t]
		\centerline{\resizebox{1\columnwidth}{!}{\includegraphics{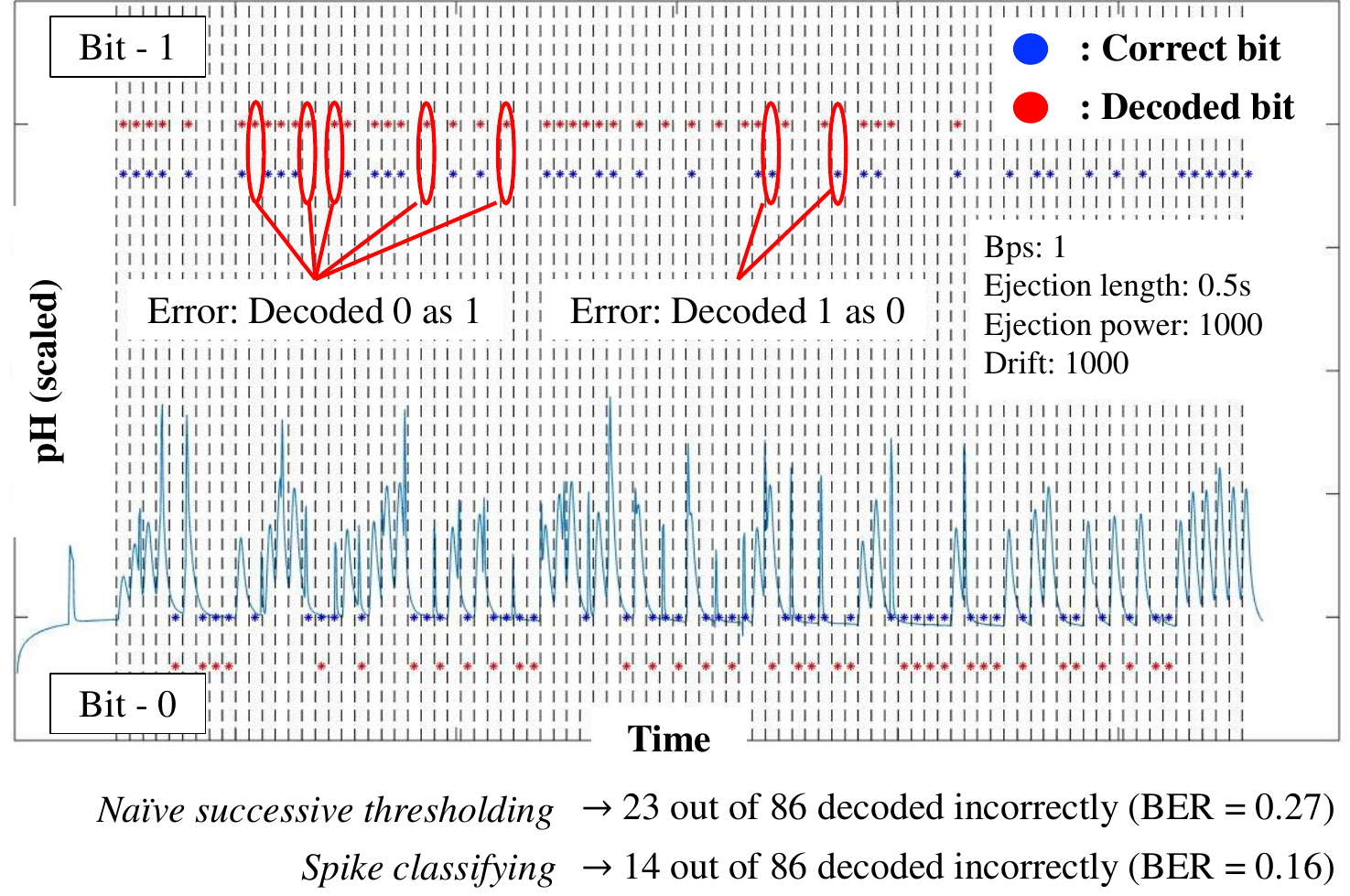}}}
		\caption{Decoding performance for a sample received signal. Compares the output of the \textit{na\"{\i}ve successive thresholding} with the \textit{spike classifying}.}
		\label{Fig.Sample_Transmission}
        \noindent\rule{\linewidth}{0.4pt}
	\end{figure}

The sample signal sequence consists of five header bits formed by the pilot signal and an additional 81 random bits. Fig.~\ref{Fig.Sample_Transmission} illustrates the decoding performance of the \textit{spike classifying} algorithm. Vertical dotted lines separate the bits in each time slot. Bits 0 and 1 are represented according to pH, red points indicate the bits decoded at the receiver, while blue points represent the bits transmitted.

	\subsubsection{Na\"{\i}ve successive thresholding}
    Each $\hat{\mathbf{x}}_i$ is determined from $\mathbf{y}_i$ by a fixed threshold. To account for chemical drift, the threshold must be compensated dynamically. We achieve this using a threshold ratio, $\eta_{\mathrm{nst}}$, which is tuned on a training dataset to minimize errors, and determined as follows:
    \begin{align}
	    \textit{threshold} = \eta_{\mathrm{nst}}\times \frac{1}{4}\sum_{j=1}^{4}\bigl(\max(\textbf{y}_j)-\textbf{y}_j(1)\bigr), \label{eq:1}\\
        \textit{current signal strength} =\max \left(\textbf{y}_i\right)-\textbf{y}_i(1), \label{eq:2}
	\end{align}
    where $\mathbf{y}_i(j)$ denotes the $j$-th element of a vector $\mathbf{y}_i$.
    
    Here, the quartile in the threshold term refers to the mean response value for the bit ‘1’, given that the first pilot bits are ({1, 1, 1, 1, 0}). The first element of the vector is included to observe the relative difference from the previous response. For each bit, the mapping method is defined as
    
    \vspace{0.7\baselineskip}
    $\text{pilot signal} =
    \begin{cases}
    1 & \text{if ~\eqref{eq:2} $>$ ~\eqref{eq:1}} \\
    0 & \text{otherwise}.
    \end{cases}$
    \vspace{0.7\baselineskip}
    
    Even with this dynamic compensation, this method produced an error rate of $26.7\%$ on sample sequence, primarily due to mechanical errors. The optimal threshold must be predeterminded before actual communication, although we manually selected the threshold that yielded the lowest average error, performance cannot be guaranteed across different sensor chips without sufficient calibration data for each one.

	\subsubsection{Spike classifying}
    The second algorithm addresses the irregular noise spikes caused by mechanical vibrations at the junction of the confluence region leading to the receiver. For minor oscillations, applying a moving‑average window to smooth the signal can mitigate their influence on adjacent regions. However, large spikes significantly affect the signal magnitude in adjacent bits, rendering simple smoothing insufficient. In such cases, the spikes are separately classified, and linear interpolation is applied after decomposing them into height and sharpness components as
    \begin{equation}
	\label{Eq.linear_intp}
	\bar{\textbf{y}}(k)=\textbf{y}(i)+\frac{k-i}{j-i}\cdot(\textbf{y}(j)-\textbf{y}(i)).
	\end{equation}
    
    Let $\mathbf{y}_{i:j}=(\mathbf{y}(i), \mathbf{y}({i+1}),\dots, \mathbf{y}({j}))$ denote the segment vector, and $\bar{\mathbf{y}}$ its corrected counterpart. If, in a sub‑interval shorter than the symbol interval, a value higher than both $\mathbf{y}(i)$ and $\mathbf{y}(j)$ occurs, the feature is defined as sharpness; conversely, the difference between the maximum value and the two ends is defined as highness. These classified values are determined empirically, and their reliability improves with additional observations; using this algorithm, we reduced the error count by nine bits compared with the previous approach.
    
	\subsubsection{Vanilla ANN}
	The last algorithm \textit{Vanilla ANN} is a multilayer perceptron that maps \textbf{y}$_i$ into $\hat{\textbf{x}}_i$ which is the purest form of artificial neural network. The multilayer perceptron is a stack of layers where each layer can approximate any continuous function with a compact domain. The composition of multiple layers is capable of emulating the more complex form of mappings with any desirable closeness given enough computation power and time~\cite{Chen1995UAT}.
	
	The layer consists of nodes with their value, and each node value is produced by multiplying a weight matrix to the input vector and pass through an activation function. A multilayer perceptron problem is an optimization problem that finds proper weight matrices and activation functions that minimize the closeness as an objective function. From extensive prevalent works on the ANN, we choose the activation function as a sigmoid function and update the weights by the Adam-Gradient algorithm which is a stochastic gradient method optimized for the neural networks starting from random initial weights. Remaining design parameter is the number of layers and the number of nodes for each layer.
	In this chapter, we employed ten nodes per each layer with three layers other than the input and the output layer.
	
	The algorithm performance showed a high dependency on the choice of training sets. A specific combination of training set provides better performances than the previous algorithms for test sets as well, while the other combination yields even worse performances. Due to insufficient data, this algorithm seems susceptible to cherry picking problem which means that the algorithm is hardly reliable for general reproductions.
	
	The input of training data handles the weights update. The number of nodes for an input layer depends on the symbol time interval; hence the sets of training are required for each case separately. By controlling the total length of input and train the system for the whole, not bit-wise, we can unify the training sets but this full training method too much resources of computational power. Moreover, a single data is a response of multiple bits hence massive measurements are required for constant training. Note that it is tough to collect extensive data with the nanomachine, this motivates the design of more efficient algorithms that can achieve better performance with limited data.

\section{Machine intelligence-aided detection}\label{sec4}
    This section describes the algorithms and logic that extend the methods introduced in Sec.~\ref{sec3} by training them through various machine-aided approaches to achieve enhanced performance. Rather than presenting precise mathematical derivations, our focus is on experimental methodologies and self-improving algorithms. Specifically, to cope with variable bit streams, we propose an algorithm based on a \textit{Vanilla recurrent neural network (RNN)} that complements the earlier, simpler calibration algorithms. To address the challenge of limited experimental data, we then introduce a virtual response generator for data augmentation and manipulation. Finally, we present a module based universal decoder that partitions the communication chain into feature-specific blocks, enabling a more sophisticated and targeted approach a global input-output mapping.
	
	\begin{figure}[t]
		\centerline{\resizebox{1\columnwidth}{!}{\includegraphics{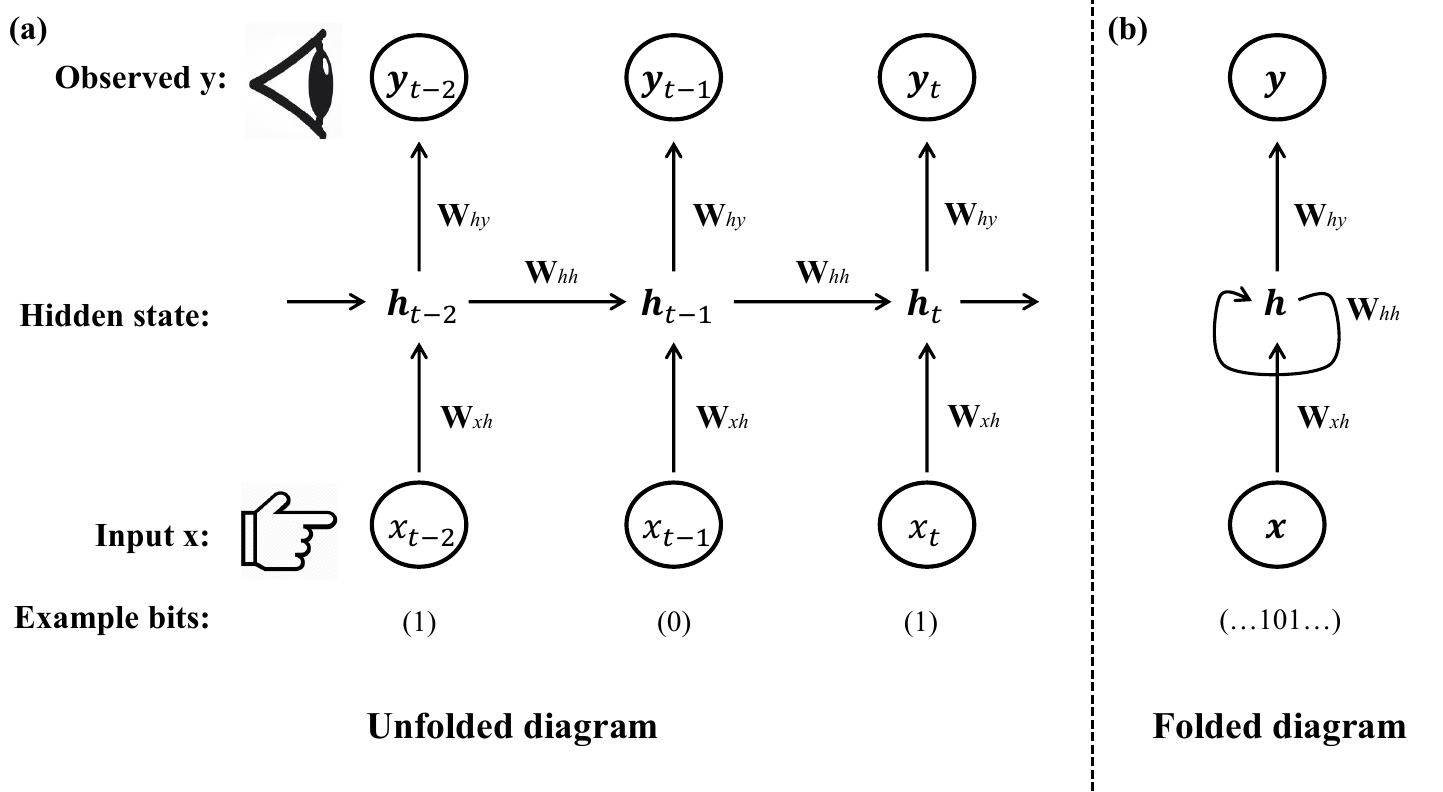}}}
		\caption{A concept diagram of how the RNN approximates sequential data. The RNN structure is interchangeably depicted with a folded version of unfolded diagram which shows the update of hidden state shares the same weight matrix. (a) An unfolded diagram. (b) A folded diagram.}
		\label{Fig.Vanilla_RNN}
        \noindent\rule{\linewidth}{0.4pt}
	\end{figure}
	
	\subsection{Vanilla RNN}
	\label{Sec4.Vanilla_RNN}

A vanilla RNN is the most fundamental recurrent neural network structure designed to model the temporal correlations of a sequential input ${x_t}^{T}$. As illustrated in Fig.~\ref{Fig.Vanilla_RNN}, it processes a sequence by sharing a single set of parameters across all time steps, making it parameter efficient. At time step $t$, the hidden state $\mathbf{h}_t$ is computed by linearly combining the input $x_t$ with the previous hidden state $\mathbf{h}_{t-1}$, followed by a nonlinear activation. The output $\mathbf{y}_t$ is then obtained from a linear transformation of $\mathbf{h}_t$:
\begin{equation}
	\label{Eq.RNN}
	\begin{split}
    \mathbf{h}_t &= \phi\bigl(\mathbf{W}_{xh}x_t + \mathbf{W}_{hh} \mathbf{h}_{t-1} + \mathbf{b}_h\bigr), \\
    \mathbf{y}_t &= \psi\bigl(\mathbf{W}_{hy} \mathbf{h}_t + \mathbf{b}_y\bigr).
	\end{split}
\end{equation}
Here, $\mathbf{W}_{xh}$, $\mathbf{W}_{hh}$, and $\mathbf{W}_{hy}$ denote the input–hidden, hidden–hidden, and hidden–output weight matrices, respectively, while $\mathbf{b}_h$ and $\mathbf{b}_y$ are bias vectors. The activation function $\phi(\cdot)$ is typically $\tanh$ or ReLU, and $\psi(\cdot)$ denotes a task-specific function. The initial hidden state $\mathbf{h}_0$ is set to the zero vector or a learnable parameter.

The network is trained using backpropagation through time (BPTT). In this process, the network is unfolded along the time axis, and gradients are propagated through the resulting multilayer perceptron. For long sequences, this process can suffer from vanishing or exploding gradients. This issue has motivated the development of techniques like gradient clipping, specialized weight initialization, and gated architectures such as long short-term memory (LSTM) and gated recurrent unit (GRU) networks.

In the proposed molecular communication system, the temporal memory of an RNN is used to model the sequential characteristics of the channel function $f$. By training the network on pairs of transmitted (\textbf{x}) and received (\textbf{y}) sequences, it can learn the forward channel model $f : \mathbf{x}\rightarrow \mathbf{y}$ or the inverse mapping for detection $f^{-1} : \mathbf{y} \rightarrow \mathbf{\hat{x}}$. The relevant approximate mapping formulas are \eqref{Eq.RNN}, and an RNN-based detector can be realized by training the network with the inputs and outputs interchanged.
    
	\begin{figure}[t]
		\centerline{\resizebox{1\columnwidth}{!}{\includegraphics{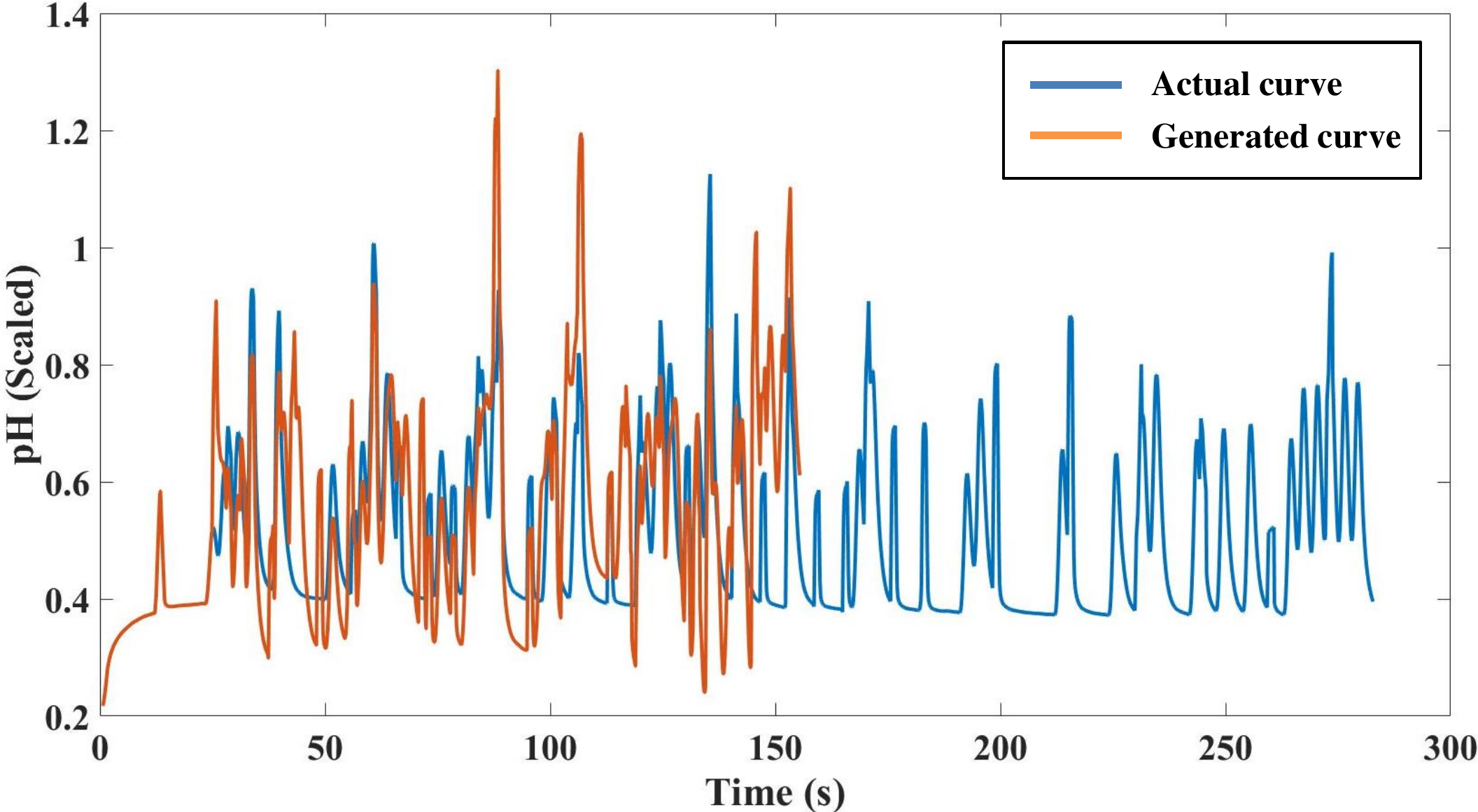}}}
		\caption{A generated response and the original response. The original curve is identical to the sample curve given in Fig. 5, and the generated curve has half the original symbol interval.}    
		\label{Fig.Curve_Manipulation}
        \noindent\rule{\linewidth}{0.4pt}
	\end{figure}
	
	\subsection{Virtual response generator}
	\label{Sec4.generative_model}
	The generation of the response curve resembles generative adversary networks (GAN)~\cite{goodfellow2014generative}. The GAN is used in machine learning for image classification and processing. In the literature, GAN consists of a generator and a discriminator where the prior produces virtual images from the trained weights of existing neural network structure while the latter discriminates the generated image between the real image and the fake image. Those two structures are an adversary to each other and by the step goes on they reciprocally develop each other.
	
	The need for virtual response generator comes from the limits of the nanomachine. The first limit is that the massive data collection requires too much time. Also, the sensors are vulnerable for long time use, massive production of the chip is not yet available, and full setup including a chamber and the measuring devices are costly albeit the chip itself is cheap. The second limit is that the data rate is not expandable due to limited specifications of the supplementary machine.
	
	The two main tasks of the virtual response generator are data augmentation and data manipulation. Data augmentation is to increase the size of the dataset where the augmented data is plausible but distinct from the others. Data manipulation is to achieve responses with manipulated environments.
	
	The generator is an approximation of the mapping $f:\mathbf{x}\rightarrow\mathbf{y}$ as~\eqref{Eq.RNN}. As described in the previous subsection, such an approximation is achievable by updating the weight matrices. While the matrices are given deterministic after the training, what we need is nonidentical generations of the virtual responses. The first idea is to add zero-mean normal random noises to given matrices. We adapted the power of noise with manual discriminations to the generated samples. The second idea is to divide the dataset into smaller groups and individually train the matrix values. We mixed the matrices from different groups to produce a new generator. 
	
	The next task we need to do is to manipulate the data rate. We can assume that the interaction between the hidden states and the input or the output is secured by fixed $\mathbf{U}$ and $\mathbf{V}$. We can rewrite~\eqref{Eq.RNN} as $h(t)=\mathcal{F}(h(t-t_s))$ where the symbol interval $t_s$ is a unit time distance of the states. The goal of data manipulation is to find $\mathcal{G}$ such that $h(t)=\mathcal{G}_k(h(t-k\cdot{t}_s))$ holds for an arbitrary positive number $k$. When $k$ is an integer, we can easily derive $\mathcal{G}_{k}=\mathcal{F}^{k}$. When $k$ is lower than 1, we need to interpolate the given function $\mathcal{F}$. For simplicity, we assume that $1/k$ is an integer. The problem is to find $\mathcal{G}_{k}$ where $\mathcal{G}_{k}^{1/k}=\mathcal{F}$ holds.
	
	We assume that $\mathcal{G}$ has the same structure with $\mathcal{F}$ as the following statement
	\begin{equation}
	\nonumber
	\exists{f_g},{W_g},{b_g} ~~s.t. ~~\mathcal{G}_k=f_g(W_gh(t-kt_s)+b_g).
	\end{equation}
	If the closed form solution is not achievable, the first approach we can try is to relax one more condition that $f_g$ is a linear activation function albeit the solution becomes suboptimal. The next approach is to apply machine learning to find solutions for the nonlinear equations.
	
	In Fig.~\ref{Fig.Curve_Manipulation} we displayed the manipulated response example with half the symbol interval; double the data rate. In the example case, we exploited the relaxation of linear relationships between the hidden states. 
    
	\subsection{Module based universal decoder}
	\label{Sec4.univ_dec_modules}

	\begin{figure}[t]
		\centerline{\resizebox{1\columnwidth}{!}{\includegraphics{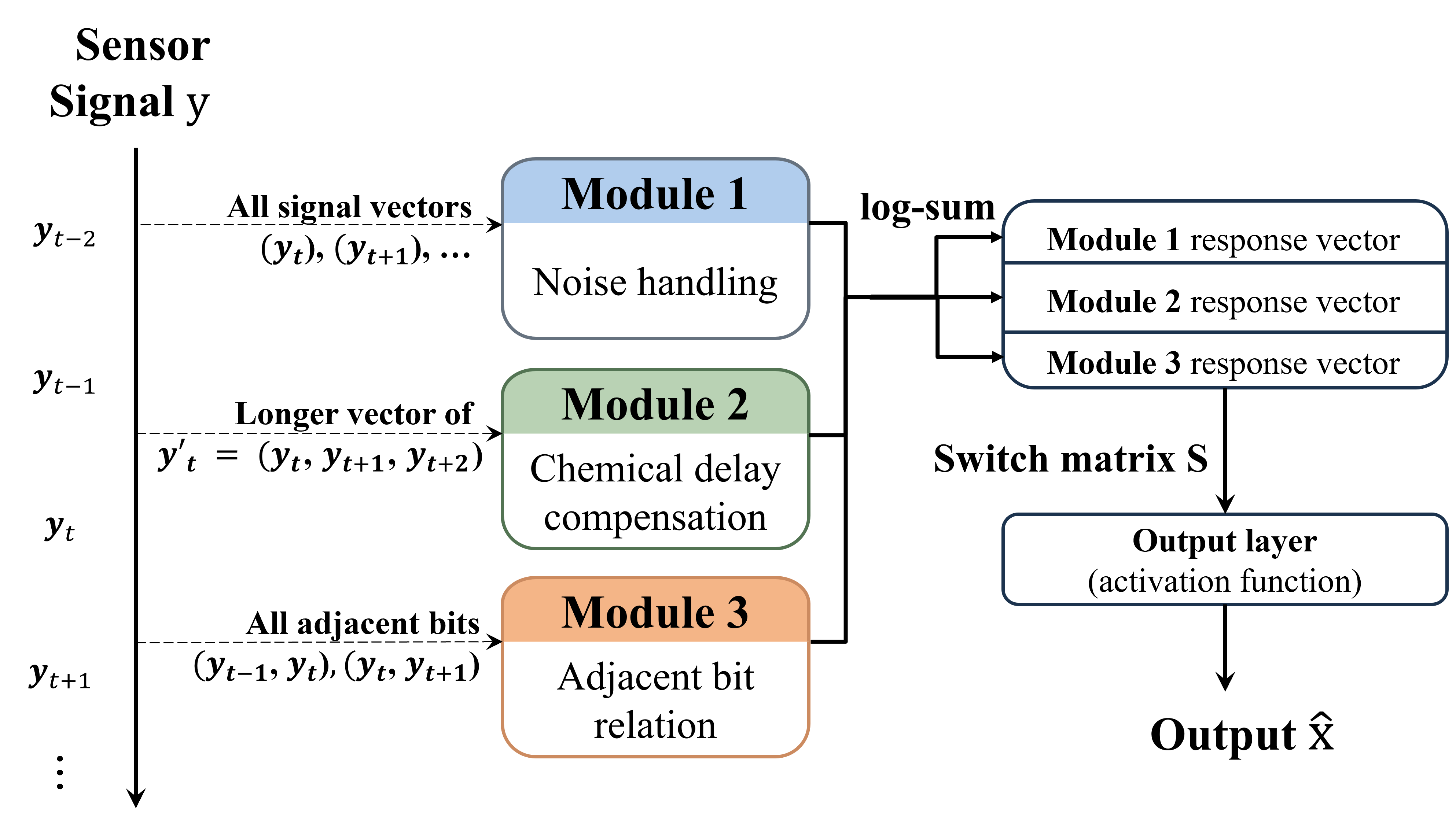}}}
		\caption{Schematic of the proposed module-based universal decoder. The architecture consists of three distinct modules that process the signal sequentially before a final decision layer merges their outputs.}
		\label{Fig.Algorithms}
        \noindent\rule{\linewidth}{0.4pt}
	\end{figure}

A standard \textit{vanilla RNN} is insufficient in this system. In addition to the error sources detailed in Sec~\ref{sec3}, there are vibration-induced noise, chemically delayed reactions caused by non-uniform concentrations on the chip surface, and single-tap ISI. These complex issues can be effectively addressed by combining the spike-classification procedure, incorporating memory elements like LSTM within the \textit{Vanilla RNN}, and analyzing bit-wise correlations. Therefore, we propose a modular decoder, as illustrated in Fig.~\ref{Fig.Algorithms}.

This approach modularizes the pipeline. Each significant error source is handled by a dedicated communication block that is optimized through separate machine learning training. This modular strategy was chosen because a monolithic, end-to-end training approach would be time-consuming, costly, and prone to poor convergence. Furthermore, excessive training on a single model risks overfitting, which undermines generalizability.

Separating the communication chain into blocks also addresses the phenomenon of the system. After a bit-1 burst is emitted by the transmitter, the chip on the receiver surface responds while simultaneous fluid flushing causes chemical delays. Consequently, noise does not accumulate indefinitely on the sensor, shortening the effective memory length relative to typical sequential data and rendering a monolithic \textit{vanilla RNN} inappropriate. Therefore, we train modules selectively, using the same switch value to activate or deactivate them as needed during the training process.

The modules are designed for (i) noise other than vibration-induced fluctuations, (ii) compensation for chemical delays, and (iii) learning the relationships between adjacent bits under single-tap memory. During initial training, all modules are enabled. Once satisfactory progress and performance are achieved, the switch values are adjusted, and when training is deemed complete, the switches are turned off. Pilot-header segments and idle intervals are distinguished so the training focuses on responses pertinent to chip behavior and learns optimal operation.

A noise handling module built upon the \textit{na\"{\i}ve successive thresholding} of Sec.~\ref{sec3}. It enhances bit-wise ANN learning by adding an activation function that outputs a confidence score for each decision and emits a vector that integrates with the other modules.

A chemical delay compensation module, using the decisions from the first module. Because OOK modulation suffers from heavy ISI after a bit-1 is transmitted, this block accounts for the number of consecutively detected bit-1s. The compensation range is adjusted according to the chain length of detected ones. To reflect the fixed-window nature of an RNN, we extend the input by feeding the concatenated vector $\mathbf{y}'_t=(\mathbf{y}_t, \mathbf{y}_{t+1}, \mathbf{y}_{t+2})$, which reduces errors by lowering the probability of false alarms in which a 1 when a 0 was transmitted.

An adjacent bit relation module using a simple ANN to learn the relation between $\mathbf{y}_t$ and $\mathbf{y}_{t+1}$ to decide whether the transmitted signal is bit 1. Except for the very first and last bits, each bit thus obtains two probability estimates, and the geometric mean is used as the final probability.

To coordinate and merge the three modules, a single additional layer combines their outputs. This layer receives the log-transformed probabilities, sums them (equivalent to multiplying the original probabilities), and applies an activation function to make a binary decision

\section{Text message transmission}\label{sec5}
In this section, we demonstrate the sample alphabet transmission via the testbed and the data detection algorithms. Each alphabet is mapped into five binary bits by the standard of International Telegraph Alphabet No.2 (ITA2).
	
	\begin{figure}[t]
		\centerline{\resizebox{1\columnwidth}{!}{\includegraphics{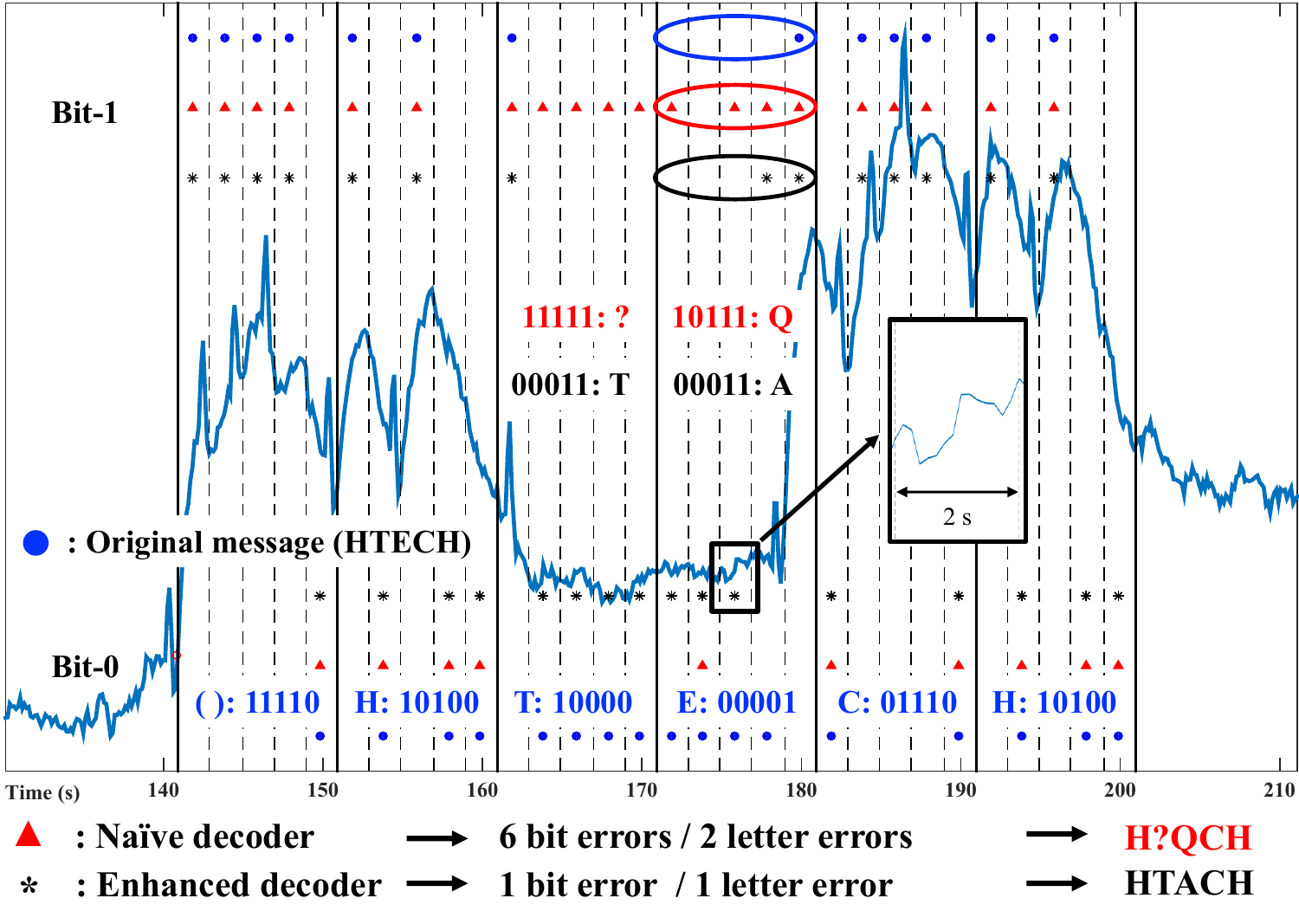}}}
		\caption{Sample letter ‘HTECH’ transmission result by 0.5 bps and decoding result with the na\"{\i}ve and the enhanced decoder.}    
		\label{Fig.Alphabet_Transmission}
        \noindent\rule{\linewidth}{0.4pt}
	\end{figure}

	In Fig.~\ref{Fig.Alphabet_Transmission}, we displayed the results of sending 'HTECH' and the detection results with \textit{na\"{\i}ve successive thresholding} and the switch-module enhanced algorithm.
	
	The result shows that the baseline algorithm yields six incorrect bit detections and two incorrect letter detections, while the enhanced algorithm results in a one-bit error that misleads the letter 'E' to 'A.' The accuracy of the transmission is 96.7\% which is higher than the average which will be shown in the next section. While the perfect detection is not securable in a given data rate, by adding an error correction code or repetition code we can enhance the accuracy as a trade-off with the data rate, to find the optimized source coding and the alphabet mapping will be our future work.
	
	While keeping the data rate and the detection accuracy, we also added features that make the communication system more versatile. The first feature is to detect the symbol interval from the pilot responses without requiring the receiver to have the prior information. We trained the machine to provide an output as one of $\{0.5, 1, 2,3\}$ for the input of the first 10s responses.
	
	To provide more pilot responses for the training, we employed the data augmentations and manipulations with the generative model. While the machine aims to find the symbol interval among the deterministic set of predefined values, the ultimate form of the machine would estimate the time difference between consecutive inputs by correctly distinguishing the noise, the spike, and the signal response. The timing difference of transmission is observed to convey higher order of data modulation with better accuracy than the concentration based modulations for a chemical receiver based communication system~\cite{Farsad2017}.

\section{Results}\label{sec6}
The experimental campaign was conducted over 14 days, with the chip replaced every day. On average, six biosensors were used in each round. Although distinct sets of biosensors were used for week 1 and week 2, the performance differences between the two sets were negligible, allowing for the aggregation of statistically meaningful result.

\begin{figure}[t]
		\centerline{\resizebox{1\columnwidth}{!}{\includegraphics{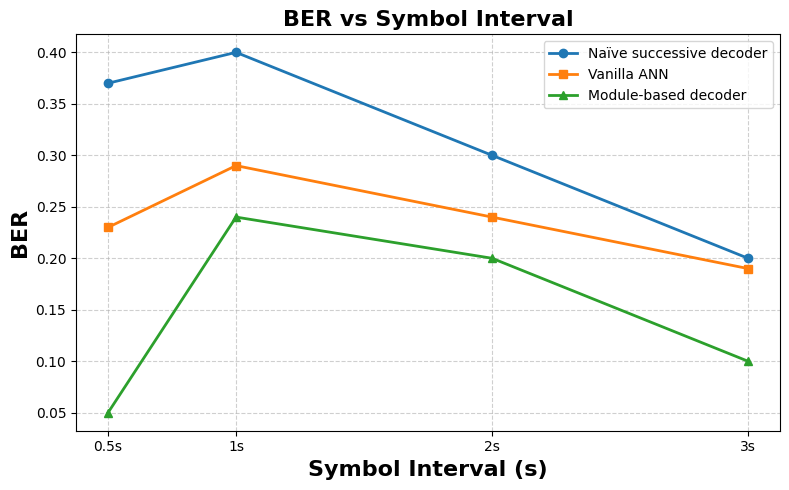}}}
		\caption{Comparison of BER performance. The graph displays the BER for the three proposed decoders across four data rates.}
		\label{Fig.Result}
        \noindent\rule{\linewidth}{0.4pt}
	\end{figure}

The experiments evaluated the three proposed algorithm using the parameters listed in Table.~\ref{Tab.Experiments_parameters}. Approximately half of the trials were performed with the flow-rate combination $(v,r)=(500,2000)\mu{\ell}/min$. Four symbol interval values were tested, $t_s\in \{ 0.5,1,2,3 \}$, and the corresponding BER are presented in Fig.~\ref{Fig.Result}. Several measurements were discarded due to sensor lifetime issues, such as rapid chip wear or excessive battery drain.

\begin{itemize}
\item \textbf{Na\"{\i}ve successive decoder:} In an OOK system, random bit guessing would yields an expected BER of 50\%. This algorithm showed limited performance, with a BER approaching 40\% at $t_s = 1\mathrm{s}$, demonstrating its limitations as a reliable decoder. Although the BER drops to about 20\% as the symbol interval lengthens, the results remain significantly lower than that of the other algorithms.
\item \textbf{Vanilla ANN:} This decoder achieved improves performance at every data rate, confirming the feasibility of communication with the nanomachine. The improvement is most pronounced for the shortest symbol interval ($t_s = 0.5\mathrm{s}$), and benefit diminishes at the longest interval. This observation implies that errors dominating in the slow-rate regime stem mainly from intrinsic testbed instabilities rather than from nonlinearities that the na\"{\i}ve decoder failed to address. Despite that instability, performance at the fastest transmission rate surpasses that of the slower cases.
\item \textbf{Module based decoder:} This method outperformed the other two across all conditions, attaining an exceptionally low BER of 5\% at $t_s = 0.5\mathrm{s}$. Interestingly, when the symbol interval was increased to 1s, the BER increased by more than a factor of four. However, consistent with the other algorithms, extending $t_s$ beyond 1s again reduces the error rate.
\end{itemize}

A common trend among all decoders is the increase in BER when the symbol interval increases from 0.5s to 1s. We attribute this behavior to a trade-off we term the \emph{washing effect}. At high data rates, the frequent emissions of liquid to transmit bit-1 also serve to periodically flush away residual molecules from the narrow channel, thereby mitigating ISI. This advantage is most pronounced at the shortest interval. For the non-modular approaches, overall performance continues to improve as $t_s$ is further lengthens, whereas the module based decoder exhibits a distinct performance optimum at the fastest rate.

\section{Conclusion}\label{sec7}

This study demonstrated a practical nanoscale molecular communication system that integrated indium–gallium–zinc-oxide (IGZO) biosensing with machine learning enhanced signal processing. We validated the concept on a complete testbed that included device fabrication, system design, and intelligent signal detection, and successfully transmitted human-readable text via chemical signaling. The biocompatible IGZO-EGFET glucose sensor represented a significant advance in hardware for nanoscale communication. Building on this foundation, our machine learning based detection pipeline substantially reduced BER more effectively than conventional methods. Within this pipeline, a virtual response generator enabled data augmentation without extensive wet-lab experiments, alleviating a critical bottleneck in molecular communication research. The modular detection pipeline addressed noise, chemical delay, and ISI while allowing targeted optimization of each impairment and maintaining end-to-end consistency. Experimental results confirmed the feasibility of nanoscale chemical text transmission. Looking ahead, this framework provides a foundation for achieving high data rates through advanced modulation, channel coding, improved synchronization, estimation, and source compression. Furthermore, it can be extended to other biochemical modalities.

\bmhead{Acknowledgements}

This work was supported by the National Research Foundation of Korea (NRF) Grant through the Ministry of Science and ICT (MSIT), Korea Government, under Grants 2022R1A5A1027646 and RS-2023-00208922.

\bibliography{sn-bibliography}

\end{document}